\documentstyle[prd,aps]{revtex}

\newcommand{\bea}{\begin{eqnarray}}
\newcommand{\eea}{\end{eqnarray}}

\begin{document}
\draft
\twocolumn[\hsize\textwidth\columnwidth\hsize\csname
@twocolumnfalse\endcsname

\title{Conserved cosmological structures in the one-loop superstring effective 
       action}
\author{Jai-chan Hwang${}^{(a)}$ and Hyerim Noh${}^{(b)}$}
\address{${}^{(a)}$ 
         Department of Astronomy and Atmospheric Sciences,
                    Kyungpook National University, Taegu, Korea \\
         ${}^{(b)}$ Korea Astronomy Observatory,
                    San 36-1, Whaam-dong, Yusung-gu, Daejon, Korea}
\date{\today}
\maketitle

\begin{abstract}

A generic form of low-energy effective action of superstring theories with
one-loop quantum correction is well known.
Based on this action we derive the complete perturbation equations
and general analytic solutions in the cosmological spacetime. 
Using the solutions we identify conserved quantities characterizing the 
perturbations: the amplitude of gravitational wave and the perturbed 
three-space curvature in the uniform-field gauge both in the 
large-scale limit, and the angular-momentum of rotational perturbation 
are {\it conserved independently of changing gravity sector}.
Implications for calculating perturbation spectra generated in the
inflation era based on the string action are presented.

\end{abstract}

\noindent

\vskip2pc]

\section{Introduction}

Superstring theory is often regarded as the leading candidate for 
unifying the gravity with the other fundamental forces and for
the quantum theory of gravity \cite{String}. 
If it is the correct theory, it may have important consequence in the
early history of our universe.
The low-energy effective action of the string theory differs from 
Einstein gravity, and the differences may leave distint 
cosmological evidences which can be probed by astronomical observations
of the cosmic microwave background radiation and the large-scale 
cosmic structures. 
The gravitational aspect of string theory can be studied using the low-energy 
effective action of string theory with loop and string tension 
($\alpha^\prime$) expansions.
Although the effective action probes the full string theory only perturbatively,
it may show some generic features of the full theory, and is expected
to be applicable in the low-energy limit before the full quantum gravitational
effect becomes important.
The generic form of the effective action of four-dimensional 
superstring model with one-loop correction is
known in the literature \cite{string-action}.
Many studies have been made on the effects of this 
action with some new results found in black hole physics and comology.
In the cosmological side, the studies are often concerned with 
possibility of realizing the non-singular universe or the inflation mechanism
\cite{cosmo-appl}.
Recent cosmological studies (based on the paradigm of 
inflation-generated-cosmic-structures) show, however, that the quantitative 
aspects of observational constraints on physics in the inflation era
are available through cosmic structure formation processes \cite{inflation}.
Results of such analyses based on the low-energy effective action 
can be found in \cite{String-pert,String-pert-axion}.

In this paper we consider the evolutions of linear stage cosmic 
structures based on sigma-model one-loop corrected string action.
We start with a general action which includes both the one-loop 
string effective action and Einstein gravity.
On a conventional cosmological model we apply the most general perturbations.
Complete sets of equations are derived and the analytic form solutions 
with important cosmological implications are found.
Based on the general solutions we identify some quantities
characterizing complete cosmic structures which remain conserved
even under the changes of the underlying gravity.
In the following we do not consider any specific cosmological scenario; 
instead, we derive general results which are applicable to any such a
scenario based on the cosmological background.
In order to make this paper self-contained we present useful
equations in the Appendix.

\section{Gravity and perturbed world model}

We consider the following general action
\bea
   S 
   &=& \int d^4 x \sqrt{-g} \Big[ {1 \over 2} f (\phi, R) 
       - {1 \over 2} \omega (\phi) \phi^{;a} \phi_{,a} - V (\phi)
   \nonumber \\
   & & - {1 \over 8} \xi (\phi) R_{GB}^2 + L_m \Big],
   \label{action}
\eea
where $f(\phi,R)$ is an algebraic function of a scalar field $\phi$
and the scalar curvature $R$, and $\omega(\phi)$, $V(\phi)$ and $\xi (\phi)$
are general functions of $\phi$;
$R_{GB}^2 \equiv R^{abcd} R_{abcd} - 4 R^{ab} R_{ab} + R^2$, and
$L_m$ is a Lagrangian of additional energy-momentum content.
The field equation and the equation of motion are presented in 
Eqs. (\ref{GFE},\ref{EOM}).
Einstein gravity with a minimally coupled scalar field is a case with 
$f=R$, $\omega = 1$, and $\xi = 0$.
The low-energy effective action of string theories is a case with
$f = e^{-\phi} R$, $\omega = - e^{-\phi}$, $V = 0$, and
with $\xi \propto e^{-\phi}$ from the one-loop string correction.
In a conformally transformed Einstein frame we have the theory with 
$f=R$, $\omega = 1$, $V = 0$, and with $\xi = \xi (\phi)$ from the one-loop 
string correction.
Equation (\ref{action}) also includes Brans-Dicke theory, non-minimally
coupled scalar field, induced gravity, $R^2$-gravity, {\it etc.}
Studies of Eq. (\ref{action}) without the $R_{GB}^2$ term have been made in
\cite{GGT-classical,GGT-quantum,GGT-application}.
Notice that our perturbation analyses in the following will be applicable 
for the general theory in Eq. (\ref{action}).
An important advantage of such a unified analysis will be explained
in \S {\it 6}.

As the metric we consider a spatially homogeneous, isotropic 
model with the most general perturbations
\bea
   d s^2 
   &=& - a^2 \left( 1 + 2 \alpha \right) d \eta^2
       - a^2 \left( \beta_{,\alpha} + B_\alpha \right) d \eta d x^\alpha
   \nonumber \\
   & & + a^2 \Big[ g^{(3)}_{\alpha\beta} \left( 1 + 2 \varphi \right)
       + 2 \gamma_{,\alpha|\beta} + 2 C_{\alpha|\beta}
   \nonumber \\
   & & + 2 C_{\alpha\beta} \Big] d x^\alpha d x^\beta,
   \label{metric}
\eea
where $a(t)$ is the cosmic scale factor with $dt \equiv a d \eta$.
$\alpha ({\bf x}, t)$, $\beta ({\bf x}, t)$, $\varphi ({\bf x}, t)$ 
and $\gamma ({\bf x}, t)$ characterize the scalar-type perturbation.
$B_\alpha ({\bf x}, t)$ and $C_\alpha ({\bf x}, t)$ are tracefree 
($B^\alpha_{\;\;|\alpha} = 0$) and correspond to the vector-type perturbation.
$C_{\alpha\beta} ({\bf x}, t)$ is transverse and tracefree
($C^\beta_{\alpha|\beta} = 0 = C^\alpha_\alpha$), and corresponds
to the tensor-type perturbation.
Indices are based on $g^{(3)}_{\alpha\beta}$ as the metric, and a vertical bar 
indicates a covariant derivative based on $g^{(3)}_{\alpha\beta}$.
We decompose the energy-momentum tensor and the scalar field as 
$T^a_b  ({\bf x}, t) = \bar T^a_b (t) + \delta T^a_b ({\bf x}, t)$
and $\phi ({\bf x}, t) = \bar \phi (t) + \delta \phi ({\bf x}, t)$; 
an overbar indicates a background order
quantity and will be omitted unless necessary.
The equations for the  background cosmological model are presented in 
Eqs. (\ref{BG1}-\ref{BG4}).
The three types of perturbations decouple from each other due to the symmetry
in the background world model and the linearity of the structures we are
assuming.
Thus, we can handle them individually.
The complete sets of equations for three perturbation types
in a spatially {\it flat} model
are presented in Eqs. (\ref{kappa}-\ref{tensor-eq}).

\section{Scalar-type perturbation}

Equations (\ref{kappa}-\ref{delta-R}) are presented in a gauge-ready form
\cite{PRW}.
Thus, we still have a right to choose one temporal gauge condition;
all variables used in the equations are spatially gauge-invariant.
Some choices are the following:
the synchronous gauge ($\alpha \equiv 0$),
the uniform-curvature gauge ($\varphi \equiv 0$),
the uniform-expansion gauge ($\kappa \equiv 0$),
the zero-shear gauge ($\chi \equiv 0$),
the uniform-field gauge ($\delta \phi \equiv 0$),
the uniform-$F$ gauge ($\delta F \equiv 0$), etc.
Except for the synchronous gauge, each one of the other gauge conditions
completely fixes the temporal gauge condition; a variable in such a gauge
condition is equivalent to a gauge-invariant combination of the
variable concerned and the variable used in the gauge condition.
A proper choice of the gauge condition is often essential for
a convenient handling of the problem which is the case in our situation.

We take the uniform-field gauge which sets $\delta \phi \equiv 0$,
thus $\delta \xi = \xi_{,\phi} \delta \phi = 0$. 
Equivalently, we can set $\delta \phi = 0$ and replace each variable with
its corresponding gauge-invariant combination with $\delta \phi$.
{}For example, $\varphi$ in the uniform-field gauge is the same as the 
gauge-invariant combination between $\varphi$ and $\delta \phi$
which is $\varphi - H \delta \phi / \dot \phi \equiv \varphi_{\delta \phi}$.
{\it Assuming} $F = F(\phi)$ we also have $\delta F = 0$.
We {\it assume} $\delta T^{(s)a}_{\;\;\;\;\; b} = 0$.
{}From Eqs. (\ref{kappa},\ref{pert2}) we can express $\alpha$ in terms of
$\dot \varphi$.
{}From Eqs. (\ref{pert1},\ref{kappa}) we can express $\kappa$ and $\chi$
in terms of $\varphi$ and $\dot \varphi$.
Thus, using either Eq. (\ref{pert3}) for $\kappa$ or Eq. (\ref{pert4})
for $\chi$ we can derive a closed form second-order differential
equation for $\varphi_{\delta \phi}$ as \cite{Kawai-attempt}
\bea
   & & {1 \over a^3 Q} \left( a^3 Q \dot \varphi_{\delta \phi} \right)^\cdot 
       - s (t) {\Delta \over a^2} \varphi_{\delta \phi} = 0,
   \label{varphi-eq}
\eea
where
\bea
   & & Q \equiv { \omega \dot \phi^2 
       + {3 \over 2} { (\dot F - H^2 \dot \xi)^2 \over F - H \dot \xi}
       \over \left( H + {1 \over 2} {\dot F - H^2 \dot \xi 
       \over F - H \dot \xi} \right)^2 }, 
   \nonumber \\
   & & s (t) \equiv 
       1 + \dot \xi { {1 \over 2} 
       \left( {\dot F - H^2 \dot \xi \over F - H \dot \xi} \right)^2
       \left( {\ddot \xi \over \dot \xi} - H - 4 \dot H 
       { F - H \dot \xi \over \dot F - H^2 \dot \xi } \right)
       \over \omega \dot \phi^2
       + {3 \over 2} { (\dot F - H^2 \dot \xi)^2 \over F - H \dot \xi} }.
   \nonumber \\
\eea
Equation (\ref{varphi-eq}) can be written in the following form
\bea
   & & \psi^{\prime\prime} + \left( s k^2 - z^{\prime\prime}/z \right) \psi = 0,
   \nonumber \\
   & & \psi \equiv z \varphi_{\delta \phi}, \quad
       z \equiv a \sqrt{ Q },
\eea
where a prime indicates a time derivative based on a conformal time $\eta$,
$a d \eta \equiv dt$, and we introduced a comoving wavenumber
using $\Delta \rightarrow - k^2$.
In the large-scale limit, $ s k^2 \ll z^{\prime\prime}/z$, thus
ignoring the Laplacian term in Eq. (\ref{varphi-eq}) we have an exact solution
\bea
   \varphi_{\delta \phi} ({\bf x}, t)
       = C ({\bf x}) + D ({\bf x}) \int_0^t
       {1 \over a^3 Q} dt, 
   \label{varphi-sol}
\eea
where $C({\bf x})$ and $D({\bf x})$ indicate the coefficients of the
growing and decaying solutions, respectively.
Thus, ignoring the transient solution (which is higher order in the
large-scale expansion compared with the solutions in the other gauges
\cite{GGT-classical}), 
$\varphi_{\delta \phi}$ is {\it conserved} in the large-scale limit.
Solutions for the other variables (even in the other gauge conditions) 
can be easily derived from our complete set of gauge-ready form equations in 
Eqs. (\ref{kappa}-\ref{delta-R}).

\section{Rotation}

{}From Eqs. (\ref{vector-eq1}-\ref{vector-eq3}), 
using notations in Eq. (\ref{vector-GI}), we have:
\bea
   & & {k^2 \over 2 a^2} \big( F - H \dot \xi \big) \Psi 
       = \left( \mu + p \right) v_\omega, 
   \nonumber \\
   & & {1 \over a^4} \big[ a^4 \left( \mu + p \right) \times
       v_\omega \big]^\cdot = - {k \over 2a} p \pi_T.
   \label{rot-eq}
\eea
If we ignore the anisotropic stress, $p \pi_T$, the angular-momentum of 
the fluid is conserved as
\bea
   a^4 \left( \mu + p \right) \times v_\omega ({\bf x}, t) \sim L({\bf x}).
   \label{rot-sol}
\eea

\section{Gravitational wave}

{}From Eq. (\ref{tensor-eq}) we have
\bea
   {1 \over a^3 Q_g} \left( a^3 Q_g \dot C^\alpha_\beta \right)^\cdot
       - s_g (t) {\Delta \over a^2} C^\alpha_\beta 
       = {1 \over Q_g} \delta T^{(t)\alpha}_{\;\;\;\;\; \beta},
   \label{GW-eq}
\eea
where 
\bea
   Q_g \equiv F - H \dot \xi, \quad
       s_g (t) \equiv {F - \ddot \xi \over F - H \dot \xi}.
\eea
Equation (\ref{GW-eq}) was studied in the string frame \cite{Gasperini}
and in the Einstein frame \cite{Kawai}.
Equation (\ref{GW-eq}) can be written in the following form
\bea
   & & \psi_g^{\prime\prime} + \left( s_g k^2 - z_g^{\prime\prime}/z_g \right) 
       \psi_g = 0,
   \nonumber \\
   & & \psi_g \equiv z_g C^\alpha_\beta, \quad
       z_g \equiv a \sqrt{Q_g}.
\eea
In the large-scale limit, $ s_g k^2 \ll z_g^{\prime\prime}/z_g$, thus
ignoring the Laplacian term in Eq. (\ref{GW-eq}), 
and ignoring the anisotropic stress, we have an exact solution
\bea
   C^\alpha_\beta ({\bf x}, t) = C^{\;\;\alpha}_{g\beta} ({\bf x})
       + D^{\;\;\alpha}_{g\beta} ({\bf x}) 
       \int_0^t {dt \over a^3 Q_g },
   \label{GW-sol}
\eea
where $C^{\;\;\alpha}_{g\beta} ({\bf x})$ and 
$D^{\;\;\alpha}_{g\beta} ({\bf x})$
indicate the coefficients of the growing and decaying solutions, respectively.
Ignoring the transient solution 
$C^\alpha_\beta$ is {\it conserved} in the large-scale limit.

\section{Discussions}

We have shown that the non-transient solutions of $\varphi_{\delta \phi}$
and $C^\alpha_\beta$ both in the large-scale limit, and the angular-momentum 
are generally {\it conserved}: see the solutions in 
Eqs. (\ref{varphi-sol},\ref{rot-sol},\ref{GW-sol}).
Remarkably, these conservation properties are valid considering 
generally time varying $V(\phi)$, $\xi (\phi)$, $\omega(\phi)$, 
and $F(\phi,R)$ [$F(\phi)$ or $f(R)$ for $\varphi_{\delta \phi}$], 
thus are valid {\it independently of changes in underlying gravity theory}.
The unified analyses of Eq. (\ref{action}) is crucially important
to make this point: that is,  
since the solutions and the conservation properties are valid considering 
general $f$, $\omega$, $V$ and $\xi$, the quantities $\varphi_{\delta \phi}$, 
$C^\alpha_\beta$ and the angular-momentum remain conserved independently 
of changing gravity sector. 
As an example, since Eq. (\ref{action}) includes both the string theory 
(possibly including the one-loop correction term) and Einstein gravity, 
the conservation properties are valid while the underlying gravity changes 
from the former to the latter one.
Indeed, this is a powerful result in the context of inflation added early 
universe scenario.
Under this scenario, the observationally relevant large-scale cosmic structures
are supposed to be generated from the quantum fluctuations (of the field
and the metric) and are pushed outside the horizon during the inflation. 
During the inflation-to-radiation transition phase the observationally
relevant scales stay in the super-horizon scale, and in such a case
our conservation properties of $\varphi_{\delta \phi}$ and $C^\alpha_\beta$ 
are applicable.
In fact, {\it it does not matter how the transition can be realized in reality:
as long as there occur transitions while the relevant scale is in 
the large-scale limit, we have the quantities conserved}.
Meanwhile, Eqs. (\ref{varphi-eq},\ref{rot-eq},\ref{GW-eq}) are
the exact equations valid in general scale.

Compared with our previous publications in
\cite{GGT-classical,GGT-quantum,GGT-application}, 
in this paper we have included the Gauss-Bonnet coupling term and have shown
that we still have similar conserved quantities. 
This extension would be particularly useful if we have an inflation scenario
where the Gauss-Bonnet coupling term has a role during the inflation.
In such a scenario, if we can calculate the quantum fluctuations (based on the
vacuum expectation values) of the metric and field variables 
when the scale is already pushed outside the Hubble horizon 
(such calculations are usually available in the many known inflation scenarios, 
see \cite{String-pert,GGT-quantum,GGT-application,NMSF}) the structures can be
described by the conserved quantities derived in the present work.
In order to make an application we need a specific inflation model
based on the action in Eq. (\ref{action}) with the $R_{GB}$ term.
Applications to specific inflation scenarios are made 
in Einstein gravity with a minimally coupled scalar field \cite{infl-pert}, 
in a non-minimally coupled scalar field \cite{NMSF}, 
in the low-energy effective action of string theory 
\cite{String-pert,GGT-application}, etc.

We expect our solutions and the general formulations made in this 
paper will be useful in probing the observationally relevant 
consequences of the superstring effective theory with one-loop correction.
In this paper, however, we have considered the roles of a dilaton field 
together with a Gauss-Bonnet type of sigma-model one-loop correction term.
We can also consider other stringy contributions from moduli fields 
and the antisymmetric tensor fields (axion) and the dual product of 
Riemann tensors \cite{string-action,cosmo-appl,choi}.
The roles of the axion coupling term, $\nu (\phi) R \tilde R$
with $R \tilde R \equiv \eta^{abcd} R_{ab}^{\;\;\;\;ef} R_{cdef}$,
have been recently considered in \cite{Choi-etal-1999}: this term vanishes
in the homogeneous-isotropic background world model, has no contributions
to the scalar- and vector-type perturbations, and the gravitational wave
is again described by conserved quantities which depend on the polarization
states.

Apparently, by considering these additional fields we will have
the multi-component situation which will especially affect the
scalar-type perturbation: in the $n$-component situation, generally, we will
have a coupled $2n$-th order differential equation and in general we do
not expect a conserved quantity in such a situation.  
However, we expect the rotation (of the fluid components) 
and the gravitational wave will not be affected by the presence of the 
additional fields in the sense that the conservation properties
of such perturbation will remain valid.
We would like to investigate the roles of these additional contributions 
in the process of cosmological structure formation in future occasions.

\section*{Acknowledgments}

We thank Profs. S. Kawai and J. Soda for drawing our attention to 
the subject and useful discussions during 19th Texas Symposium in Paris.
We also wish to thank Profs. K. Choi, R. Easther, M. Gasperini and P. Kanti
for useful discussions and comments.

\section*{Appendix}
\setcounter{equation}{0}
\def\theequation{A\arabic{equation}}

{\it A. The field equation and the equation of motion:}
{}From variations of Eq. (\ref{action}) we have:
\bea
   & & F G^a_b = \omega \Big( \phi^{;a} \phi_{,b}
       - {1 \over 2} \delta^a_b \phi^{;c} \phi_{,c} \Big)
       - {1 \over 2} \delta^a_b \left( RF - f + 2 V \right)
   \nonumber \\
   & & \quad
       + F^{;a}_{\;\;\;\; b} - \delta^a_b \Box F
       + \left( R^a_{\;\;cbd} - R_{cd} \delta^a_b
       + R_{cb} \delta^a_d \right) \xi^{;cd}
   \nonumber \\
   & & \quad
       + G^a_c \xi^{;c}_{\;\;\; b} - G^a_b \Box \xi 
       + T^a_b,
   \label{GFE} \\
   & & \Box \phi + {1 \over 2 \omega} \Big( \omega_{,\phi} \phi^{;a} \phi_{,a}
       + f_{,\phi} - 2 V_{,\phi} - {1 \over 4} \xi_{,\phi} R_{GB}^2 \Big) = 0,
   \nonumber \\
   \label{EOM}
\eea
where $F \equiv {\partial f \over \partial R}$, and
$T_{ab}$ is an additional energy momentum tensor defined as
$\delta ( \sqrt{-g} L_m ) \equiv {1 \over 2} \sqrt{-g} T^{ab} \delta g_{ab}$.

\vskip .3cm
{\it B. Background equations:} 
{}From Eqs. (\ref{GFE},\ref{EOM},\ref{metric}) and $T^b_{0;b} = 0$ we have:
\bea
   & & H^2 + {K \over a^2} = {1 \over F - H \dot \xi} \Big[ {1 \over 6} 
       \left( \omega \dot \phi^2 + R F - f + 2 V \right) 
   \nonumber \\
   & & \quad
       - H \dot F - {1 \over 3} T^0_0 \Big], 
   \label{BG1} \\
   & & \dot H - {K \over a^2} = {1 \over 2(F - H \dot \xi)} 
       \Big[ - \omega \dot \phi^2
       - \ddot F + H \dot F 
   \nonumber \\
   & & \quad
       + \left( \ddot \xi - H \dot \xi \right)
       \Big( H^2 + {K \over a^2} \Big)
       + T^0_0 - {1 \over 3} T^\alpha_\alpha \Big], 
   \label{BG2} \\
   & & \ddot \phi + 3 H \dot \phi + {1 \over 2 \omega}
       \Big[ \omega_{,\phi} \dot \phi^2 - f_{,\phi} + 2 V_{,\phi}
   \nonumber \\
   & & \quad
       + 6 \xi_{,\phi} \left( \dot H + H^2 \right) 
       \Big( H^2 + {K \over a^2} \Big) \Big] = 0,
   \label{BG3} \\
   & & \dot T^0_0 + 3 H T^0_0 = H T^\alpha_\alpha,
   \label{BG4}
\eea
where an overdot indicates the time derivative based on $t$ and $K$ 
is the sign of three-space curvature;
$H \equiv {\dot a \over a}$ and $R = 6 ( \dot H + 2 H^2 + K/a^2 )$. 
Equation (\ref{BG3}) also follows from Eqs. (\ref{BG1},\ref{BG2},\ref{BG4}).

\vskip .3cm
{\it C. Perturbation equations:}
{}From Eqs. (\ref{GFE},\ref{EOM},\ref{metric}) we can derive the following 
set of equations (we assume a spatially flat model, thus $K = 0$ and 
$g^{(3)}_{\alpha\beta} = \delta_{\alpha\beta}$):
\bea
   & & \kappa \equiv - 3 \left( \dot \varphi - H \alpha \right)
       - {\Delta \over a^2} \chi,
   \label{kappa} \\
   & & \delta T^{(s)0}_{\;\;\;\;\; 0} = 2 \left( F - H \dot \xi \right)
       \Big( H \kappa + {\Delta \over a^2} \varphi \Big)
       - \omega \dot \phi^2 \alpha 
   \nonumber \\
   & & \quad
       + \left( \dot F - H^2 \dot \xi \right)
       \left( \kappa + 3 H \alpha \right)
       + \omega \dot \phi \delta \dot \phi 
   \nonumber \\
   & & \quad
       + {1 \over 2} \left( \omega_{,\phi} \dot \phi^2 - f_{,\phi}
       + 2 V_{,\phi} \right) \delta \phi
       - 3 H \delta \dot F
   \nonumber \\
   & & \quad
       + \Big( 3 \dot H + 3 H^2 + {\Delta \over a^2} \Big) \delta F
       + H^2 \Big( 3 H \delta \dot \xi - {\Delta \over a^2} \delta \xi \Big),
   \label{pert1} \\
   & & \delta T^{(s)0}_{\;\;\;\;\; \alpha} 
       = {1 \over a} \Big[ - {2 \over 3} \left( F - H \dot \xi \right)
       \Big( \kappa + {\Delta \over a^2} \chi \Big)
       - \left( \dot F - H^2 \dot \xi \right) \alpha
   \nonumber \\
   & & \quad
       + \omega \dot \phi \delta \phi + \delta \dot F - H \delta F
       - H^2 \left( \delta \dot \xi - H \delta \xi \right) \Big]_{,\alpha},
   \label{pert2} \\
   & & \delta T^{(s)\gamma}_{\;\;\;\;\; \gamma} - \delta T^{(s)0}_{\;\;\;\;\; 0}
   \nonumber \\
   & & \quad
       = 2 \left( F - H \dot \xi \right) \Big( \dot \kappa + 2 H \kappa
       + 3 \dot H \alpha + {\Delta \over a^2} \alpha \Big)
   \nonumber \\
   & & \quad
       + \left[ \dot F + H^2 \dot \xi - 2 ( H \dot \xi)^\cdot \right]
       \left( \kappa + 3 H \alpha \right)
       + 3 \left( \dot F - H^2 \dot \xi \right) \dot \alpha
   \nonumber \\
   & & \quad
       + 2 \left( 2 \omega \dot \phi^2 + 3 \ddot F - 3 H^3 \dot \xi \right)
       \alpha 
       - 2 \left( \ddot \xi - H \dot \xi \right) {\Delta \over a^2} \varphi
   \nonumber \\
   & & \quad
       - 4 \omega \dot \phi \delta \dot \phi
       - \left( 2 \omega_{,\phi} \dot \phi^2 + f_{,\phi} 
       - 2 V_{,\phi} \right) \delta \phi
   \nonumber \\
   & & \quad
       - 3 \delta \ddot F - 3 H \delta \dot F 
       + \Big( 6 H^2 + {\Delta \over a^2} \Big) \delta F 
       + 3 H^2 \delta \ddot \xi 
   \nonumber \\
   & & \quad
       + \left( 2 \dot H + H^2 \right)
       \Big( 3 H \delta \dot \xi - {\Delta \over a^2} \delta \xi \Big),
   \label{pert3} \\
   & & \delta T^{(s)\alpha}_{\;\;\;\;\; \beta}
       - {1 \over 3} \delta T^{(s)\gamma}_{\;\;\;\;\; \gamma}
       \delta^\alpha_\beta
       = - {1 \over a^2} \Big( \nabla^\alpha \nabla_\beta
       - {1 \over 3} \delta^\alpha_\beta \Delta \Big)
   \nonumber \\
   & & \quad
       \Big[ \left( F - H \dot \xi \right) \left( \varphi + \alpha
       - \dot \chi - H \chi \right)
       - \left( F - H \dot \xi \right)^\cdot \chi 
   \nonumber \\
   & & \quad
       - \left( \ddot \xi - H \dot \xi \right) \varphi
       + \delta F -\left( \dot H + H^2 \right) \delta \xi \Big],
   \label{pert4} \\
   & & \delta \dot T^{(s)0}_{\;\;\;\;\;0} 
       + H \left( 3 \delta T^{(s)0}_{\;\;\;\;\;0} 
       - \delta T^{(s)\alpha}_{\;\;\;\;\;\alpha} \right)
       + \left( 3 T^0_0 - T^\alpha_\alpha \right) \dot \varphi
   \nonumber \\
   & & \quad
       - {1 \over a} \delta T^{(s)0|\alpha}_{\;\;\;\;\;\alpha}
       + {1 \over a^2} \left( T^0_0 \Delta \chi
       - T^\alpha_\beta \chi^{|\beta}_{\;\;\;\alpha} \right) = 0,
   \label{pert5} \\
   & & \dot T^{(s)0}_{\;\;\;\;\;\alpha} 
       + 4 H T^{(s)0}_{\;\;\;\;\;\alpha} 
       + {1 \over a} T^{(s)\beta}_{\;\;\;\;\;\alpha,\beta}
       + {1 \over a} \left( \alpha + 3 \varphi \right)_{,\beta} T^\beta_\alpha
   \nonumber \\
   & & \quad
       - {1 \over a} \left( T^0_0 \alpha + T^\beta_\beta \varphi 
       \right)_{,\alpha} = 0,
   \label{pert6} \\
   & & \delta \ddot \phi + \left( 3 H + {\omega_{,\phi} \over \omega}
       \dot \phi \right) \delta \dot \phi
       + \Big[ \left( {\omega_{,\phi} \over \omega} \right)_{,\phi}
       {\dot \phi^2 \over 2} 
   \nonumber \\
   & & \quad
       + \Big( {- f_{,\phi} + 2 V_{,\phi} \over 2 \omega} \Big)_{,\phi} 
       + 3 H^2 \left( \dot H + H^2 \right) \Big( {\xi_{,\phi} \over \omega}
       \Big)_{,\phi} 
   \nonumber \\
   & & \quad
       - {\Delta \over a^2} \Big] \delta \phi
       = \dot \phi \left( \kappa + \dot \alpha \right) 
       + \left( 2 \ddot \phi + 3 H \dot \phi + {\omega_{,\phi} \over \omega}
       \dot \phi^2 \right) \alpha 
   \nonumber \\
   & & \quad
       + {1 \over 2 \omega} \left( F_{,\phi} - H^2 \xi_{,\phi} \right) \delta R
       + {2 \over \omega} \xi_{,\phi} \dot H \Big( H \kappa
       + {\Delta \over a^2} \varphi \Big),
   \label{pert7} \\
   & & \delta R = 
       - 2 \Big[ \dot \kappa + 4 H \kappa + 3 \dot H \alpha
       + {\Delta \over a^2} \left( 2 \varphi + \alpha \right) \Big]; 
   \nonumber \\
   & & \quad
       \delta R_{GB}^2 = 
       4 H^2 \delta R - 16 \dot H \Big( \kappa + {\Delta \over a^2}
       \varphi \Big), 
   \label{delta-R} \\
   & & \delta T^{(v)0}_{\;\;\;\;\; \alpha} = - {1 \over 2} 
       \left( F - H \dot \xi \right)
       {\Delta \over a^2} \left( B_\alpha + a \dot C_\alpha \right),
   \label{vector-eq1} \\
   & & \delta T^{(v)\alpha}_{\;\;\;\;\; \beta} = {1 \over 2 a^3}
       \Big\{ a^2 \left( F - H \dot \xi \right) 
       \Big[ \left( B_\beta^{\;\;|\alpha} + B^\alpha_{\;\;,\beta} \right)
   \nonumber \\
   & & \quad
       + a \left( C_\beta^{\;\;|\alpha} + C^\alpha_{\;\;,\beta} 
       \right)^\cdot \Big] \Big\}^\cdot,
   \label{vector-eq2} \\
   & & \dot T^{(v)0}_{\;\;\;\;\;\alpha} + 4 H T^{(v)0}_{\;\;\;\;\;\alpha} 
       + {1 \over a} \delta T^{(v)\beta}_{\;\;\;\;\;\alpha,\beta} = 0,
   \label{vector-eq3} \\
   & & \delta T^{(t)\alpha}_{\;\;\;\; \beta}
       = (F - H \dot \xi) \ddot C^\alpha_\beta 
   \nonumber \\
   & & \quad
       + \left[ (F - H \dot \xi)^\cdot + 3 H (F - H \dot \xi) \right] 
       \dot C^\alpha_\beta
   \nonumber \\
   & & \quad
       - ( F - \ddot \xi ) {\Delta \over a^2} C^\alpha_\beta,
   \label{tensor-eq}
\eea
where $\Delta$ is a Laplacian based on $\delta_{\alpha\beta}$.
Equations (\ref{kappa}-\ref{delta-R}), Eqs. (\ref{vector-eq1}-\ref{vector-eq3}),
and Eq. (\ref{tensor-eq}) completely describe the scalar-, vector-, and
tensor-type perturbations, respectively;
we decomposed the perturbed energy-momentum tensor as
$\delta T^a_b = \delta T^{(s)a}_{\;\;\;\;\; b}
+ \delta T^{(v)a}_{\;\;\;\;\; b} + \delta T^{(t)a}_{\;\;\;\;\; b}$,
where superscripts $(s)$, $(v)$, and $(t)$ indicate the three
perturbation types.
Equation (\ref{pert5},\ref{pert6},\ref{vector-eq3}) follow from 
$T^b_{a;b} = 0$.
Some useful quantities for deriving the equations can be found in our study
of $R^{ab} R_{ab}$ gravity in \cite{Rab}.

{}For the scalar-type perturbation we introduced a spatially gauge-invariant 
combination $\chi \equiv a (\beta + a \dot \gamma )$. 
In the above set of equations we have not fixed any gauge condition;
thus the equations are written in a gauge-ready form.
Equation (\ref{pert7}) also follows from 
Eqs. (\ref{kappa}-\ref{pert3},\ref{pert5}) and Eq. (\ref{BG2}).
{}For the vector-type perturbation we introduce the following variables:
\bea
   & & T^{(v)0}_{\;\;\;\;\;\alpha} \equiv (\mu + p) v_\omega Y_\alpha, \quad
       \delta T^{(v)\alpha}_{\;\;\;\;\;\beta} \equiv p \pi_T Y^\alpha_\beta, 
   \nonumber \\
   & & B_\alpha + a \dot C_\alpha \equiv \Psi Y_\alpha; \quad
       \Delta Y_\alpha \equiv - k^2 Y_\alpha, 
   \nonumber \\
   & & Y_{\alpha\beta} \equiv - {1 \over 2 k} ( Y_{\alpha|\beta}
       + Y_{\beta|\alpha} ),
   \label{vector-GI}
\eea
where $\mu \equiv - T^0_0$ and 
$p \equiv {1 \over 3} T^\gamma_\gamma$;
$Y_\alpha$ is a vector-type ($Y^\alpha_{\;\;\;|\alpha} = 0$) harmonic function
with a wavenumber $k$.



\begin{references}
\bibitem{Kawai-attempt}
         The authors of \cite{Kawai} tried to analyse the scalar-type
         equations in Einstein frame without success.
         {}In order to derive this equation, our choice of the uniform-field 
         gauge is essentially important to make the analyses simple.
\end{references}
\end{document}